\documentclass[letterpaper]{article} 
\usepackage{aaai19}  
\usepackage{times}  
\usepackage{helvet}  
\usepackage{courier}  
\usepackage{url}  
\usepackage{graphicx}  
\usepackage{amsmath}
\usepackage{amsfonts}
\usepackage{xcolor}
\usepackage{tabularx}
\begin{document}
\title{Powerful, transferable representations for molecules through intelligent task selection in deep multitask networks}
\author{
  Clyde Fare\\
  IBM Research UK \\
  Sci-Tech Daresbury\\
  Warrington, UK. \\
\And 
  Lukas Turcani\\
  IBM Research UK \\
  Sci-Tech Daresbury\\
  Warrington, UK. \\
\And
  Edward O.~Pyzer-Knapp\thanks{Corresponding author} \\
  IBM Research UK \\
  Sci-Tech Daresbury\\
  Warrington, UK. \\
  \texttt{epyzerk3@uk.ibm.com} \\}
\maketitle
\begin{abstract}
Chemical representations derived from deep learning are emerging as a powerful tool in areas such as drug discovery and materials innovation. Currently, this methodology has three major limitations - the cost of representation generation, risk of inherited bias, and the requirement for large amounts of data. We propose the use of multi-task learning in tandem with transfer learning to address these limitations directly. In order to avoid introducing unknown bias into multi-task learning through the task selection itself, we calculate task similarity through pairwise task affinity, and use this measure to programmatically select tasks. We test this methodology on several real-world data sets to demonstrate its potential for execution in complex and low-data environments. Finally, we utilise the task similarity to further probe the expressiveness of the learned representation through a comparison to a commonly used cheminformatics fingerprint, and show that the deep representation is able to capture more expressive task-based information. 
\end{abstract}

\section{Introduction}

Representations within chemical informatics are a currently active area of research with many neural fingerprint architectures emerging recently \cite{Schutt2018}, \cite{Duvenaud2015}, \cite{Kearnes2016}, \cite{Gilmer2017}. These representations while powerful, have three major drawbacks:
\begin{enumerate}
    \item They require large quantities of labelled data.
    \item They are expensive to train.
    \item They do not generalise well outside of the task for which they were initially trained.
\end{enumerate}

Fortunately, as with image and speech recognition, sources of high quality data for chemical inference have significantly increased over the past decades - see e.g. \cite{Gaulton2017}, \cite{Hachmann2011b}, \cite{Lopez2016} however, the availability of data is not homogeneously spread over all chemical tasks of interest. Additionally, data sets which do exist are often small - for example the majority of datasets on http://cheminformatics.org/datasets/ are below 1000 compounds. This limits the direct applicability of neural representations which typically require larger datasets. As a consequence, more traditional non-neural fingerprints such as the extended connectivity fingerprints (ECFP) are still commonly used in industry.  These representations have had some successes as representations for deep learning see e.g. \cite{Unterthiner2014a} and \cite{pyzer2015learning}, but have some significant disadvantages. These fingerprints contain bias since they were constructed for particular tasks (in the case of ECFP, QSAR modelling for pharmaceutical molecules).  Within deep learning, multitask networks offer several advantages for generation of a data driven representation including data augmentation, attention focusing, eavesdropping, biasing the representation towards greater generality and regularisation \cite{Ruder2017}.  Once a sufficiently general information dense is built, it is possible to make use of the representation within a target inference task. Thus training a model with far fewer parameters than a full neural fingerprint architecture and hence requiring far less labelled data, whilst potentially achieving similar levels of performance.

When building a representation, it is important to fully consider which tasks to build a representation from. There had been some indication that, within drug discovery in particular, multitask learning acts as a panacea and simply increasing the number of tasks improves performance\cite{Ramsundar2015}.  More recent analyses, however,  have shown that negative transfer also affects drug discovery; with tasks that are negatively correlated hindering performance \cite{Xu2017}. Thus, in the face of a large number of available chemical tasks, choosing the right support tasks to match a particular chemical inference task of choice remains a key challenge. 

Recent work in image recognition \cite{Zamir2018} investigated the taxonomy of various image tasks and used the resulting task dictionary to aid transfer learning. In this work, a similar approach is taken to problems in chemical inference. Rather than focus on finding combinations of single task representations that are of utility to each other, as in \cite{Zamir2018}, here the focus is on how to choose support tasks that a multitask representation will be directly learned from, such that the final representation will prove useful for other holdout tasks. Since the majority of the computational costs is spent on the construction of a representation, holdout tasks are fast to train, thus avoiding the problem of computational cost with neural fingerprints. 

In this study, we make use of a novel combination of the Weave neural fingerprint modified to use a set-to-set network for generation of the final fixed length representation.  We train a representation using selected support tasks, and compare against the commonly used ECFP fingerprint \cite{Rogers2010} on a set of holdout tasks, as implemented in RDKIT \cite{Landrum2006}. 

\section{Contribution}
The contributions of this paper are:

\begin{itemize}
    \item Introduction of a neural fingerprint architecture that implicitly obeys physical symmetries without information discarding steps.
    \item Development of a framework for selection of support tasks for training neural fingerprints within a multitask network that will be suitable for particular target tasks.
    \item Evaluation of the framework for choosing support tasks for five target chemical inference tasks.
\end{itemize}


\section{Related Work}



Neural fingerprints have become increasing popular following the introduction of graph convolution networks to chemical inference problems \cite{Duvenaud2015}. Duvenaud created a differentiable version of the classic ECFP fingerprint which could be learned via backpropagation. This was expanded upon by Kearnes who separated the graph network structure from the atom identities \cite{Kearnes2016}. Both Kearnes and Duvenaud showed promising performance though their neural fingerprints did not outperform ECFPs in all cases. Several other graph convolution networks have been introduced that act on explicit molecular coordinates e.g. \cite{Schutt2018} rather than over the molecular graph though here we restrict our attention to those that act on the molecular graph alone.

 Transfer learning from multitask networks within chemical inference task tasks has seen limited investigation with Ramsundar \cite{Ramsundar2015} finding that if transfer learning to non training tasks was possible it required large amounts data. 
 
 Given the proliferation of hard weight sharing multitask networks, consideration of how to choose tasks has long been considered with typical advice to choose 'similar' tasks \cite{Caruana1998}, though the notion of  similarity is still subject to investigation. In the NLP setting Alonso \cite{Alonso2016} and Ruder \cite{Ruder2017b} have suggested that beneficial tasks had compact uniform label distributions while Bingel has suggested features extracted from single task learning curves could be used to predict multitask performance within dual task networks \cite{Bingel2017}. 

Our approach to task choice is inspired by Zamir \cite{Zamir2018}, who built a taxonomy of image tasks and used them to transfer learn representations from multiple single tasks. They showed that by using their task taxonomy to identify useful tasks they could reduce the amount of labelled data needed to solve target tasks by approximately 2/3rds at minimal performance loss.

\section{Problem setup}
\subsection{Requirements of a molecular representation}
In order to generate a reasonable representation for molecules, it is necessary to take into account scientific principles which determine molecular construction. This has the additional advantage than inspection of the representation is likely to be interpretable. 

We place the following constraints on a molecular representation:
\begin{enumerate}
    \item The order in which atoms are labelled does not change the representation (i.e. the representation must be invariant to atom ordering)
    \item The order in which the bonds are labelled does not change the representation (i.e. the representation must be invariant to bond ordering)
    \item The way in which pairs within the molecule are labelled should not change the representation (i.e. the representation must be invariant to pair labelling).
\end{enumerate}

\subsection{Multi-task Representation learning}
Suppose we have some inference task(s) that correspond to pairs of molecular input data $X$ and targets $y$ where $X \in \mathbb{R}^{N*D}$, N is the number of samples and D is the dimensionality of each sample and $y \in R \ or \ [0,1]$ the former for regression tasks and the latter for binary classification (that could be generalised to $\mathbb{Z}$ for multinomial classification).
We wish to find an approximation of the function that maps $X$ to $y$, i.e. $\hat{y}=f(X,\lambda)$. We do so by optimising some parametric function (in our case a deep network) parameterised by parameters $\lambda$ with optimal parameters $\lambda^*$. To find $\lambda^*$ we minimise an appropriate loss function (in our case either the mean square error or binary cross entropy) :
$$\lambda^* = \underset{\lambda}{\arg\min{L(X,\lambda )}}$$

We decompose $f$ into the product of a representation function $f_r$ that maps the input data into latent space and a tuning function $f_t$ that in turn maps from the latent space to the target space.  
$f(X,\lambda) = f_t(f_r(X, \lambda_r), \lambda_t )$ 

We then suppose that we can approximate the ideal $\lambda{_r}^*$ for some particular target task $X_t$ by optimising it over some other set of support tasks $X_i \in {S}$. This assumes that there is some underlying similarity between the target tasks and the support tasks and in particular that there exists some representation function that extracts features from the input domain that are useful across the multiple different tasks.

We define a loss over the support tasks $L_S$ such that all support tasks share the same $\lambda_r$ but whose $\lambda_t$ are allowed to differ:

$$\lambda{_r}^*,\lambda{_t}* = \underset{\lambda_r, \lambda_t}\arg\min{L_i(X_i, \lambda_r, \lambda_{ti})}$$

This corresponds the classic hard weight sharing multitask network, we then make use of $\lambda{_r}^*$ to generate representations for use with our target tasks reducing the number parameters needed to be learned for the target inference tasks to $\lambda_t$.


\section{Methods}
\subsection{Architecture}
\begin{figure}
\begin{minipage}[c]{0.48\textwidth}
\centering
  \includegraphics[width=0.95\linewidth]{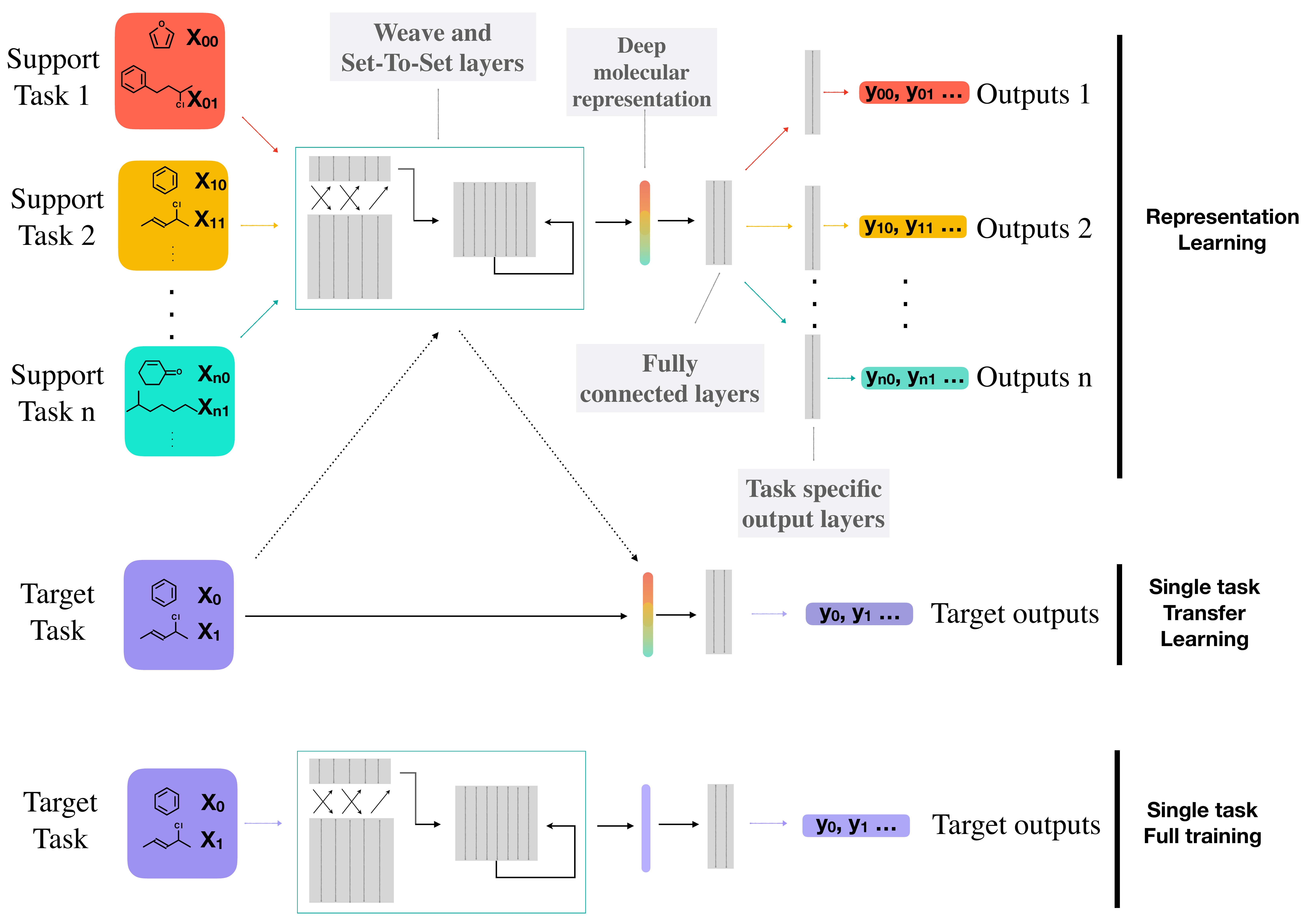}
  \caption{Schematic of the architectures used to train the deep molecular representation (top). For transfer learning experiments (middle) and for single task experiments (bottom). Components used are weave graph convolution blocks, a set-to-set block, a fully connect block and for the multitask training a final task specific linear output layer. Dashed lines to and from the weave and set-to-set blocks for the transfer learning network indicate the weights are frozen while training.\label{arch}}
\end{minipage}
\end{figure}

The architecture introduced in this paper uses a combination of the weave graph convolution network \cite{Kearnes2016} to extract features from the molecular graph and a set-to-set architecture \cite{Vinyals2015} to create a fixed length shared representation. Finally, fully connected layers were attached to the representation using ReLu activation with a last linear layer to generate the output. See Figure \ref{arch} for depictions of the multitask, single task and transfer learned networks used. 

A brief description of the weave and set-to-set networks now follows. (For full account see \cite{Kearnes2016} and \cite{Vinyals2015} respectively).

\subsection{Input data structure}

The architecture takes as input two layers; an atomic layer and a pair layer. These encode the identity of the atoms with each molecule and how they are connected to one another within the molecular graphs respectively.

The atomic layer corresponds to a vector of atom identities for each molecule where each identity corresponds to a one hot encoding that classifies whether the atom is one of the light main group elements (Hydrogen, Carbon, Nitrogen, Oxygen, Fluorine, Phosphorus, Sulphur, Chlorine, Bromine, Iodine) or a metal atom.

The pair layer corresponds to a matrix of pair identities for each molecule where each  identity corresponds to a one hot encoding that classifies the type of bond present between the two atoms in the pair (none, single, partial double, double, triple) and an encoding of the graph-distance (i.e. the fewest edges needed to be traversed to reach one atom node from the other). 

\subsection{Weave}

 We use the weave convolution approach detailed in \cite{Kearnes2016}, which works as follows: the atom and pair layers are initially passed through separate convolution layers yielding independently transformed atom and pair layers but subsequent convolutions applied to the pair layers take inputs from the transformed atom layers and vice versa. This 'weaving' between the two input sources gives rise to the name of the architecture and aims to allow patterns to be extracted that depend on both the identity of the atoms as well as their connectivity. After some number of passes the transformed vector of atomic identifies is taken as the output. The fully convolutional architecture can accept molecules of different sizes but in order to generate a fixed length representation for the final stage of inference extra steps must be taken. In the weave architecture simply summing or binning of the variable-length representation is performed in order to achieve this. This step discards information however it allows the final fixed length representation to be invariant to the order of the atomic indices an important aspect for molecular inference as physical principles demand that the order the atoms are specified in cannot influence molecular properties.

\subsection{Set-to-set}

As an alternative to the information discarding final step of the weave architecture we use a set to set architecture as described in \cite{Vinyals2015}. This is capable of learning sequence to sequence transformations that are invariant to the order the sequence is supplied in making it suitable for tasks operating sequences for which no order is natural. It makes use of an LSTM with attention where the set to be learned is read into the memory of the LSTM, the LSTM is then run with neither inputs nor outputs but updating the memory for some number of steps before finally generating the fixed length output sequence. As the attention mechanism sums over all memory elements the output is necessarily invariant to the input order.

\subsection{Task choice}

When building representations which involve more than one task, a subtle form of bias (or uncertainty) can emerge from the particular choice of support tasks.  This can be thought of as conceptually similar to bias-variance trade off in general model building.  It is possible to select a narrow range of similar tasks which will build a powerful local representation which is unlikely to generalize, or a wide rage of dissimilar tasks, which are unlikely to have strong local performance.  Either of these situations can be tolerated, and indeed desirable, for a given problem, so long as the location on this task bias-variance scale is known. It is thus important to choose the support tasks in such as way that this uncertainty is understood.  

We propose a framework for choice of support tasks making use of a pairwise score matrix computed over the library of available support tasks. Having computed this score matrix we select the $n$ support tasks that have maximal scores when paired with the target tasks where $n$ is the number support tasks we are choosing to use. The pairwise scores correspond to the improvement (or lack thereof) of a task whilst being co-trained in a dual task network over that task being trained within a single task network. Thus elements of the score matrix are defined as

$$S_{ij} = L^{DT}_{ij} - L^{ST}_{i}$$ where $S$ is the score matrix, $L^{ST}_i$ is the final loss of the trained single task network trained on the $i$th task and evaluated on holdout data from the $i$th task, $L^{DT}_{ij}$ is the final loss of the trained dual task network trained on tasks $i$ and $j$ and evaluated on holdout data from the $i$th task.

Whilst this method does not take into account higher order effects, which we believe have an increasing effect as the number of tasks increases, we believe that it captures significant information to be of use, and future work will detail an extension to this method for increasingly large numbers of support tasks.

\subsection{Training details}
All tasks were trained using TensorFlow 1.5 \cite{Abadi2016} as built on PowerAI \cite{Cho2017} and run on Power 8+ systems utilising P100 GPUS. In order to account for different sizes of dataset, and remove potential artifacts derived from overlapping sample sets, a maximum of 1000 randomly selected samples per task were extracted. An overview of the tasks are provided in table \ref{tab:task_groups} while a complete list is present in the supporting information. For each task, 20\% of the data was used as the evaluation set. A single set of hyperparameters was derived from exploration of single-task performance, and were used throughout - see table \ref{tab:hyperparams}. Batch normalisation was used within the weave and fully connected layers and dropout was used within the fully connect layers. Optimisation of the weight matrix during training made use of the Adam optimizer. \cite{Kingma2015}.

\begin{table}
\caption {WSTS hyperparamers} \label{tab:hyperparams} 
\begin{center}
 \begin{tabular}{|l c|} 
 \hline
Parameter & Value \\ [0.5ex] 
 \hline
Weave blocks & 2 \\
Weave block neurones & \begin{tabular}{@{}c@{}}24x24x24 \\ 24x24x24\end{tabular} \\ 
RNN timesteps & 8 \\
Fully connected neurones & 1000x100 \\
Final representation size & 256 \\
Learning rate & $1*10^{-5}$ \\
Dropout & 30\% \\
Batch size & 100\\
\hline 
\end{tabular}
\end{center}
\end{table}

\section{Experiments}

\subsection{Pairwise task scores}

\begin{figure}
\centering
  \includegraphics[width=\linewidth]{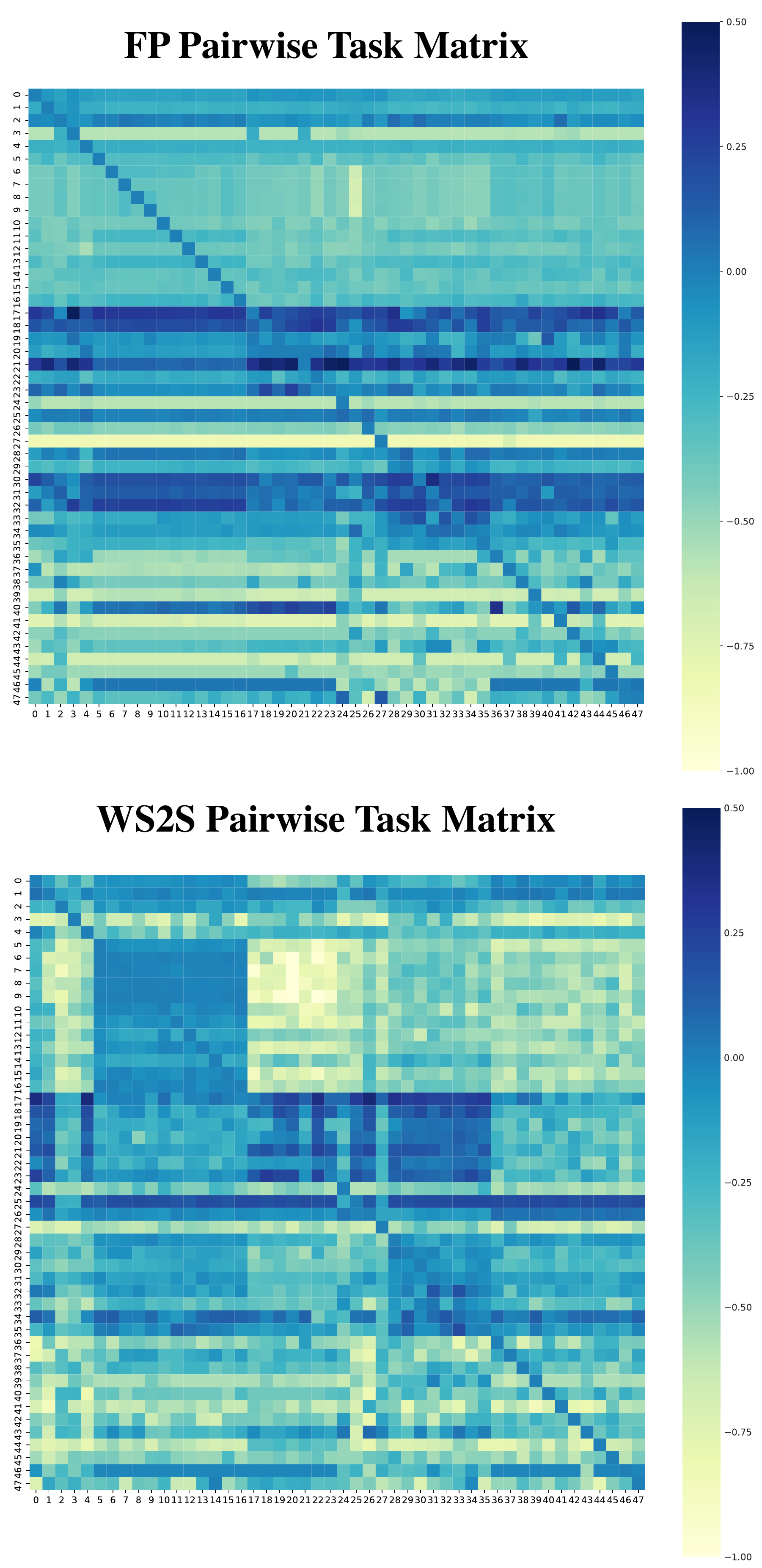}
  \caption{Matrices of pairwise task improvements for 48 chemical tasks using a simple Morgan finger print based architecture (FP) on the left and using the weave/set-to-set architecture (WSTS) on the right. The magnitude of each pixel corresponds to the improvement (or decline) of a task when it is trained within a dual task network with a partner task. Improvement is relative to single task training with the same architecture. Rows specify the primary task being considered whilst columns specify the partner tasks. Tasks are specified by ID, and arranged in order of the dataset the tasks were extracted from. See Table \ref{tab:task_groups} for an overview. \label{score_M}}
\end{figure}

\begin{table}
\caption {Task IDs} \label{tab:task_groups} 
\begin{center}
 \begin{tabular}{|l c|} 
 \hline
Dataset Description & Task ids \\ [0.5ex] 
 \hline
Experimental solubilities & 0 \\
Melting points & 1 \\
Anti-HIV data & 2 \\
Blood-brain-barrier data & 3 \\
Power conversion efficiencies & 4 \\
Ab-initio derived properties & 5-16 \\
Experimental optical properties & 17-23 \\ 
Anti-malarial properties & 24-25 \\
Octonal-water partition data & 26 \\
Microsome stability & 27 \\
Experimental thermochemical properties & 28-35 \\
Toxicity data & 36-47 \\
\hline
\end{tabular}
\end{center}
\end{table}

Figure \ref{score_M} shows the pairwise scores computed for all pairs of the 48 tasks contained  within our task library both for our WSTS architecture and for a simplified multitask architecture using Morgan fingerprints (FP) as the representation and correspondingly only training the fully connected layers. Switching the output of the graph convolution and set-to-set layers with Morgan fingerprints but keeping the fully connected layers and task specific output layers the same allows us to compare the effect of co-training on the individual tasks with two very different choices of representation. The effect of of multitasking on drug discovery task performance was characterised by Xu et al. \cite{Xu2017} as showing improvement if the molecules were 'structurally similar' and their targets were correlated, showing regression if they were 'structurally similar' and their targets anti-correlated and making no difference if the input molecules were 'structurally dissimilar'. Given that structurally similarity implicitly depends on the representation and these two networks make use of very different representations it might be expected that there would be no similarity between the two score matrices yet this is not the case. 

The influence of the thermochemical data on the other tasks is broadly similar across both architectures (columns 28-35), showing greater benefit to the optical properties (rows 17-23), to the other thermochemical proeprties (rows 28-35) and to some extend the solubilities, melting points and octonol-water coefficient (rows 0,1,26) compared the biological datasets (rows 2,3,36-47). The basic physical properties benefiting from co-training with one another more so than with biological tasks is chemically plausible as the biological tasks are more likely to benefit from features that express steric interactions with large molecules which will be less relevant for basic physical properties. Whilst comparing across the architectures there is more column similarity (how a given task alters the performance of other tasks) than row similarity (how a given task's performance is altered by the presence of other tasks). However, the broad trends in the row similarity are present in both with some tasks appearing promiscuous and likely to benefit from co-training and other appearing anti-social and favour solo training. This trade off may reflect underlying similarity within the tasks though further investigation along these lines is needed.

It is interesting to note, that whilst similar, the WSTS architecture shows far greater diversity compared to the Morgan FP architecture. This difference is particular pronounced regarding the influence of co-training with the ab-initio tasks, where the WSTS shows benefits for co-training from other ab-initio tasks and negative transfer when co-trained with the experimental optical properties. This pattern is much weaker if present at all for the Morgan FP data. We believe that the greater representational flexibility afforded by the neural fingerprints allow them to capture a far richer task information landscape. Some further examples of this are the solubility data (row 0) showing greater benefit in the WSTS case from the octanol-water partition data (column 26) vs. the thermochemical and  optical data. This is consistent with physical reasoning as the preference for octonol vs. water is known to be very strongly determined by solubility. Similarly the positive effect of the toxicity data (columns 36-47) and to lesser effect the anti HIV activity (column 2) and anti-malarial data (columns 24-25) on prediction of the solubility is present to a much degree within the WSTS data than the Morgan FP data and is physically intuitive given biological activity is highly dependent on solubility. A further physically plausible similarity present within both datasets is the similar behaviour of the ab-initio formation energies (rows 6-9). These are very strongly correlated properties and the corresponding rows of the score matrix are similarly strongly correlated. Not all physical expectations are realised however - e.g. co-training the computed molecular formation energies and their experimental counter parts in the thermochemical dataset (tasks 6-9 vs. tasks 31-32) is less beneficial than might have been anticipated.

\begin{figure}[h!!]
\centering
  \includegraphics[width=0.8\linewidth]{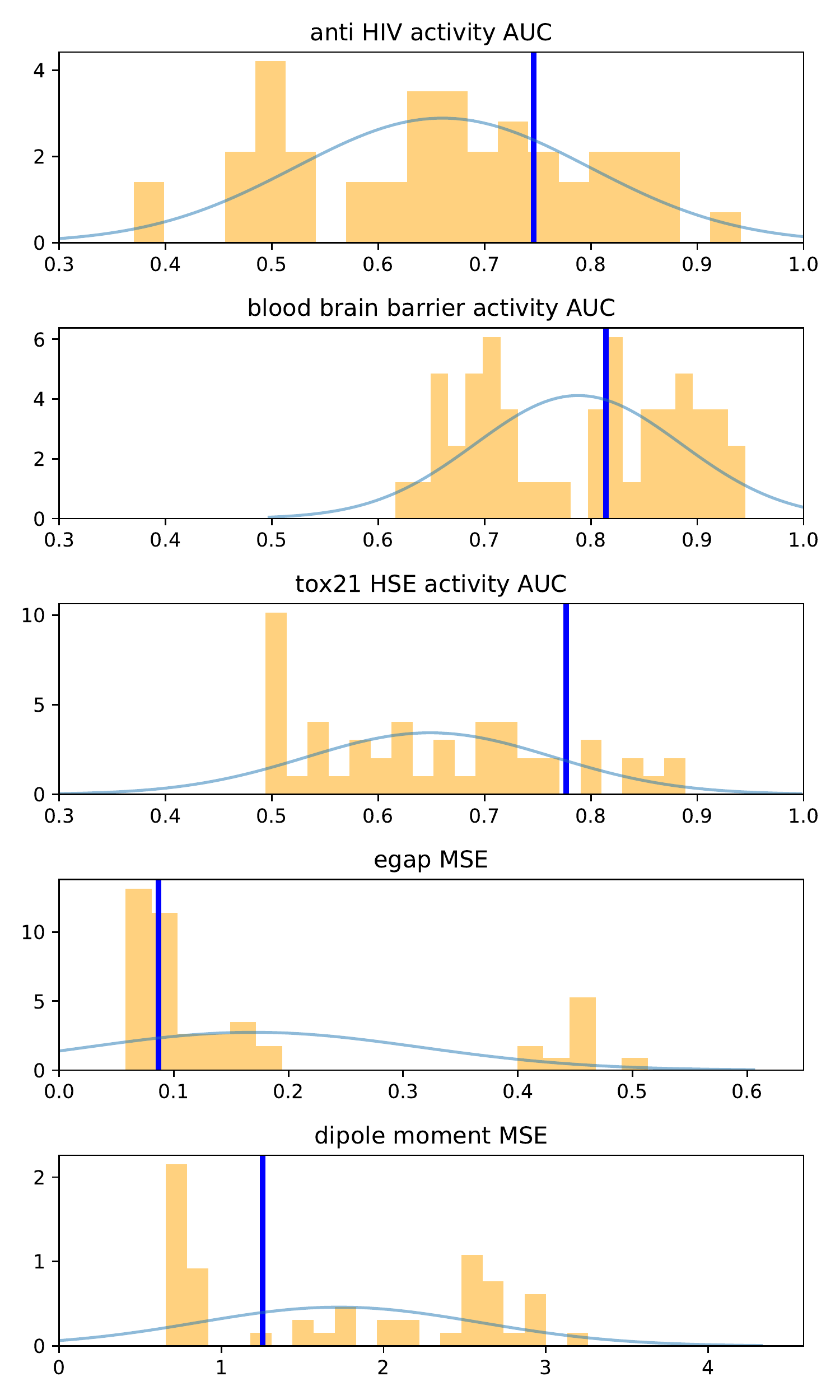}
  \caption{Transfer learning performance evaluated for 5 target tasks. Representations are trained on 4 task multitask networks that do not include the target tasks. The background histogram and fitted normal distribution represent randomly chosen support tasks, whilst the blue vertical bar represents the median result of using the pairwise task score matrix to select support tasks.  Note the top three tasks are classification tasks for which AUC is reported whilst the bottom two are regression tasks for which MSE is reported.\label{framework_perf}}
\end{figure}
\subsection{Transfer learning}
Whilst powerful, existing neural fingerprinting techniques for chemistry require re-training for each new task which is discovered; requiring significant amounts of data.  Not only is this not how the learning process happens with humans, it is also impractical in chemistry - since it is likely that the representation is being built for predicting a property of which there is little known data.  Alternatively, with the growth in use of generative models for molecular discovery \cite{Gomez-Bombarelli2018a}, it might be desired that existing data not be used for building a representation as this will affect the kinds of chemistry which the generative model will be capable of creating.  Transfer learning represents a solution to both of these problems, as a representation can be built ahead of time on large amounts of trustworthy chemical data with only a small portion of the model needing to be trained to provide significant predictive power, thus limiting the complexity of the model.  

\begin{table}[h!]
\caption {Average training time of single inference tasks using Morgan fingerprints, WSTS fingerprints and transfer learned WSTS fingerprints over the task library.} \label{tab:timings} 
\begin{center}
 \begin{tabular}{|l c |} 
 \hline
  & Training time /$s$  \\ 
  \hline\hline
  Morgan FP & $407 \pm 92$ \\
 WSTS FP & $2633 \pm 73$ \\
 Transfer WSTS FP & $2137 \pm 95$ \\
 \hline
\end{tabular}
\end{center}
\end{table}

To examine the potential of the WSTS architecture for transfer learning 5 holdout target tasks were chosen. These were composed of two regression tasks taken from materials discovery problems; an experimental band gap, and dipole moment (tasks 17 and 36), and three classification tasks taken from pharmaceutical problems; anti HIV activity, blood brain barrier permitivity and heat-shock factor response (tasks 2, 3 and 44).

\begin{table}[h!!]
\caption {Performance of Morgan fingerprints, deep molecular features, and transfer learned deep molecular features on selected tasks. Highest performance shown in bold.} \label{tab:magic_results} 
\begin{center}
\resizebox{\columnwidth}{!}{
 \begin{tabular}{ccccc} 
 \hline
 Task & Category & Morgan&WSTS & WSTS  \\ 
 & & FP &FP&Transfer FP\\
 \hline\hline
 Energy gap & Regression (MSE) & 0.690 & 0.766 & \textbf{0.077} \\ 

 Dipole moment & Regression (MSE) & 0.792 & \textbf{0.506} & 0.684 \\

 Anti HIV activity & Classification (AUC) & 0.782 & 0.954 & \textbf{0.959} \\

 Blood brain & Classification (AUC) &0.775 & \textbf{0.924} & 0.900 \\
   barrier permitivity &&&&\\

Heat shock & Classification (AUC) & 0.749 & \textbf{0.811} & 0.799 \\ 
  factor activity &&&&\\
\hline
\end{tabular}%
}
\end{center}
\end{table}

A multitask network was trained using four co-tasks (which were not allow to include the target tasks), after training the weights for the weave and set-to-set components were frozen and used within a single task network trained on the target task where only the fully connected layers were trained. Table \ref{tab:magic_results} shows the mean square error (MSE) and the area under the receiver operating characteristic curve (AUC) for the two regression tasks and the three classification tasks respectively for a simple Morgan FP single task network with only the fully connected layers trained, a full WSTS network where all weights are trained and the transfer learned WSTS network where again only the fully connected layers are trained. The four tasks used to train the representation were chosen by picking those tasks that showed greatest benefit from being co-trained with the target tasks (i.e those columns within the score matrix that had the highest value within the target task rows). We note that the transfer learned representation out performs the Morgan fingerprints in each case and in the case of the energy gap regression task and the anti HIV classification task even out performs the fully trained WSTS network. 

To indicate the uncertainty which can permeate into representations through task selection, we compared representations generated from 4 task networks with 4 randomly chosen support tasks (again restricted to not include the target task) with the aforementioned representations trained using a task network using the optimal target tasks given the pairwise scores as discussed above. For each target task, 10 sets of 4 randomly chosen support tasks were co-trained 5 times using 5 different seeds for a total of 50 generated representations per target task. These representations were then used within a single task WSTS network trained on the target task with only the fully connected layers optimised. Figure \ref{framework_perf} shows the performance of transfer learning from randomly chosen tasks verses the median performance for the set of tasks chosen using the pairwise task scores. Here, the median was selected over the mean due to the robustness of this statistic against outliers, in small sample size situations and to avoid over optimism. It can be seen that actively selecting tasks never hurts the expected performance, and in the majority of the cases studied, provided a performance boost over the expected value derived from random task selection. Figure \ref{framework_perf} also demonstrates the uncertainty which can be derived from task selection.  Whilst there do exist some seeds of some of the randomly built representations which give superior performance, in all cases the range of potential performance is significant, and could be disruptive and misleading to critical workflows in (e.g.) drug discovery and materials innovation.  
Timings for training a single task using the Morgan fingerprints the WSTS architecture and the transfer learned WSTS fingerprints can be seen in table \ref{tab:timings}. It should be noted that we report training using frozen weights to perform the transfer learning, and thus the timings reported here are for the whole workflow including training the support tasks. In practice, one would generate the representation separately and correspondingly would achieve training times on a time scale similar to, or less than, the Morgan fingerprints depending on the size of representation generated.

\section{Conclusion}

We have described a new architecture combining the weave and set to set architectures for building powerful and flexible representations for a variety of chemical tasks.  Whilst we show that for most of the hold out tasks, a single task representation provides the best performance, the transferred representation approaches or exceeds this performance, for a fraction of the computational cost.  
In order to reduce the risk of bias in the transferred representation arising from the choice of support tasks, we propose a systematic method for choosing tasks which reduces the variablility and thus uncertainty associated with generating representations built using multiple tasks.  Finally, we note that the task distances used to select support tasks captures underlying physiochemical properties of the tasks themselves, and thus shows promise for the holy grain of an explainable, end-to-end learned representation for chemistry.


\bibliography{aaai19}
\bibliographystyle{aaai}

\end{document}